\documentclass[10pt,prc,aps]{revtex4}   
\usepackage{graphicx}
\tighten
\begin{document}
\draft
\title{ The splitting of the one-body potential \\                                                     
in spin-polarized isospin-symmetric nuclear matter 
 }              
\author{            
 Francesca Sammarruca}      
\affiliation{                 
 Physics Department, University of Idaho, Moscow, ID 83844-0903, U.S.A  } 
\date{\today} 
\email{fsammarr@uidaho.edu}
\begin{abstract}
Spin-polarized symmetric nuclear matter is studied within the Dirac-Brueckner-Hartree-Fock 
approach. We pay particular attention to the difference between the                        
one-body potentials of upward and downward polarized nucleons. This is formally analogous to the ``Lane potential" for isospin-asymmetric
nuclear matter. We point out the necessity for additional information on this fundamentally important quantity 
and suggest ways to constrain it. 

\end{abstract}
\maketitle

\section{Introduction} 
                                                                     
Describing 
the properties of spin and isospin {\it symmetric} nuclear matter still presents considerable intellectual 
challenges. For instance, the physical pictures of the underlying one-particle fields are very different 
in relativistic and non-relativistic approaches. In relativistic models, saturation mechanisms are introduced
through negative energy Dirac states, whereas non-relativistic approaches must be implemented with three-body
forces (TBF) in order to describe saturation properties correctly. Although the relation between the two      
philosophies seems to be understood in terms of TBF of the ``Z-diagram" type,                                   
saturation details can be quite different 
in the two frameworks. 

When other aspects are considered, such as spin and/or isospin polarized states of nuclear matter, conclusions     
become even more model dependent and available constraints are very limited.
The magnetic properties of neutron/nuclear matter have been studied
extensively with a variety of
theoretical methods [1-28]. 
In a previous calculation \cite{SK07}, we have investigated spin-polarized pure neutron matter (NM). Such system has  gathered much attention lately, in relation to the issue of possible ferromagnetic instabilities.        
Also, the possibility of strong magnetic fields in the interior 
of neutron stars makes the study of polarized NM important and timely. 

Although these are very exciting issues, there are other motivations for studies of polarized matter.
Here, for instance, we will focus on the spin degrees of freedom of symmetric nuclear matter (SNM), having in mind 
a terrestrial scenario as a possible ``laboratory". 
We will pay particular attention to the spin-dependent {\it symmetry potential}, namely the gradient between               
the single-nucleon potentials for upward and downward polarized nucleons. 
The interest around this quantity arises because of its natural interpretation as a 
spin dependent nuclear optical potential, defined in perfect formal analogy with the Lane potential \cite{Lane} for the isospin degree
of freedom in isospin-asymmetric nuclear matter (IANM). 

Concerning optical potential analyses, to the best of our knowledge, spin degrees of freedom have not been given much attention,
possibly due to the increased difficulties in obtaining empirical constraints as compared to the unpolarized system. 
Another way to access information related to the spin dependence of the nuclear interaction in nuclear matter
is the study of collective modes such as spin giant resonances (SGR). However,
those are not easily observed with sufficient strength \cite{pol33,Oster92}. 

In summary, if constraints on the isospin-dependent properties of the nuclear equation of state (EoS) are
still scarce, those on the spin dependence are even more so, and should be pursued along with 
theoretical predictions of the spin-dependent
nucleon field.                                   
What makes this issue particulary interesting is that spin degrees of freedom and relativity 
are inherently tied with each other. Thus, comparison between relativistic and non-relativistic
predictions should be insightful.

This paper is organized as follows: In the next Section, we review the main aspects of the formalism.
We will then demonstrate the splitting of the one-body potential in spin-asymmetric matter 
and discuss its significance. 
Our conclusions are summarized in the last Section. 

\section{Formalism } 

Our calculation is microscopic and treats the nucleons
relativistically.                                                  
Within the Dirac-Brueckner-Hartree-Fock (DBHF) method, 
the interactions of the nucleons with the nuclear medium are expressed as self-energy corrections to the 
nucleon propagator. That is, the nucleons are regarded as ``dressed" quasi-particles.                                     
Relativistic effects lead to an intrinsically density-dependent interaction which is consistent 
with the contribution from TBF typically employed in non-relativistic approaches.
The advantage of the DBHF approximation is the absence of phenomenological TBF to be extrapolated
at higher densities from their values determined through observables at normal density.

The starting point of any microscopic calculation of nuclear
structure or reactions is a realistic free-space nucleon-nucleon interaction. A
realistic and quantitative model for the nuclear force with
reasonable foundations in theory is the one-boson-exchange (OBE)
model \cite{Mac89}. Our standard
framework consists of the Bonn B potential together with the
DBHF approach to nuclear matter. A
detailed description of our application of the DBHF method to
nuclear, neutron, and asymmetric matter can be found in a recent review of our 
work \cite{FS10}.

Similarly to what we have done to describe isospin asymmetries of
nuclear matter, the single-particle potential is the solution of a
set of coupled equations, 
\begin{equation}
U_u = U_{ud} + U_{uu}
\end{equation}
\begin{equation}
U_d = U_{du} + U_{dd} \; , 
\end{equation}
to be solved self-consistently along with the two-nucleon $G$-matrix.                   
In the above equations, $u$ and $d$ refer to spin-up and spin-down polarizations,
respectively, and each $U_{\sigma \sigma '}$ term contains the
appropriate (spin-dependent) part of the interaction, $G_{\sigma
\sigma '}$. More specifically,
\begin{equation}
U_{\sigma}({\vec p}) = \sum _{\sigma '=u,d} \sum _{q\leq k_F^{\sigma
'}} <\sigma,\sigma '|G({\vec p},{\vec q})|\sigma,\sigma '>,
\end{equation}
where the second summation indicates integration over the Fermi
sea of spin-up (or spin-down) nucleons, and
\begin{widetext}
\begin{eqnarray}
<\sigma,\sigma '|G({\vec p},{\vec q})|\sigma,\sigma '>&=&
\sum_{L,L',S,J,M,M_L} <\frac{1}{2} \sigma;\frac{1}{2} \sigma '|S
(\sigma + \sigma ')>
<\frac{1}{2} \sigma;\frac{1}{2} \sigma '|S (\sigma + \sigma ')> \nonumber\\
&\times&<L M_L;S(\sigma + \sigma ')|JM>
<L' M_L;S(\sigma + \sigma ')|JM> \nonumber\\
&\times& i^{L'-L} Y^{*}_{L',M_L}({\hat k_{rel}}) Y_{L,M_L}({\hat
k_{rel}}) <LSJ|G(k_{rel},K_{c.m.})|L'SJ> \; . 
\end{eqnarray}
\end{widetext}
The notation $<j_1m_1;j_2m_2|j_3m_3>$ is used for the Clebsh-Gordan
coefficients. Clearly, the need to separate the interaction by spin
components brings along angular dependence, with the result that the
single-particle potential depends also on the direction of the
momentum. The $G$-matrix equation is solved using
partial wave decomposition and the matrix elements are then summed
as in Eq.~(4) to provide the new matrix elements in the
uncoupled-spin representation needed for Eq.~(3). Furthermore, the
scattering equation is solved using relative and center-of-mass
coordinates, $k_{rel}$ and $K_{c.m.}$, which are then easily related
to the momenta of the two particles, $p$ and $q$, in order to
perform the integration indicated in Eq.~(3).  Notice that solving
the $G$-matrix equation requires knowledge of the single-particle
potential, which in turn requires knowledge of the interaction.
Hence, Eqs.~(1-2) together with the $G$-matrix equation constitute a
self-consistency problem, which is handled, technically, exactly the
same way as previously done for the case of isospin asymmetry
\cite{FS10,AS1}. The Pauli operator for scattering of two particles with
unequal Fermi momenta, contained in the kernel of the $G$-matrix
equation, is also defined in perfect analogy with the
isospin-asymmetric one \cite{AS1},
\begin{equation}
Q_{\sigma \sigma '}(p,q,k_F^{\sigma},k_F^{\sigma '})=\left\{
\begin{array}{l l}
1 & \quad \mbox{if $p>k_F^{\sigma}$ and  $q>k_F^{\sigma '}$}\\
0 & \quad \mbox{otherwise.}
\end{array}
\right.
\end{equation}
The Pauli operator is then expressed in terms of $k_{rel}$ and
$K_{c.m.}$ and angle-averaged in the usual way.

\section{The splitting of the one-body potential in spin-polarized nuclear matter}                                                                  
Figure 1 displays the average potential energy of nucleons, $<U_{u/d}>$, in polarized SNM 
as a function of the degree of spin asymmetry, described by the spin-asymmetry parameter
$\alpha = \frac{\rho_{u} - \rho_{d}} 
{\rho_{u} + \rho_{d}}$ ($\alpha>0$, allowing the spin-up species to increase in density).           
The splitting becomes more pronounced with increasing density, compare 
the left and right panels in the figure.                                                                                
As mentioned earlier, these potentials become
direction-dependent in the presence of spin asymmetry, although we found such dependence (on the polar
angle $\theta$) to be very mild. In Fig.~1, the potentials are averaged with respect to both magnitude and
direction of the momenta. 

\begin{figure}[!t] 
\centering 
\scalebox{0.41}{\includegraphics{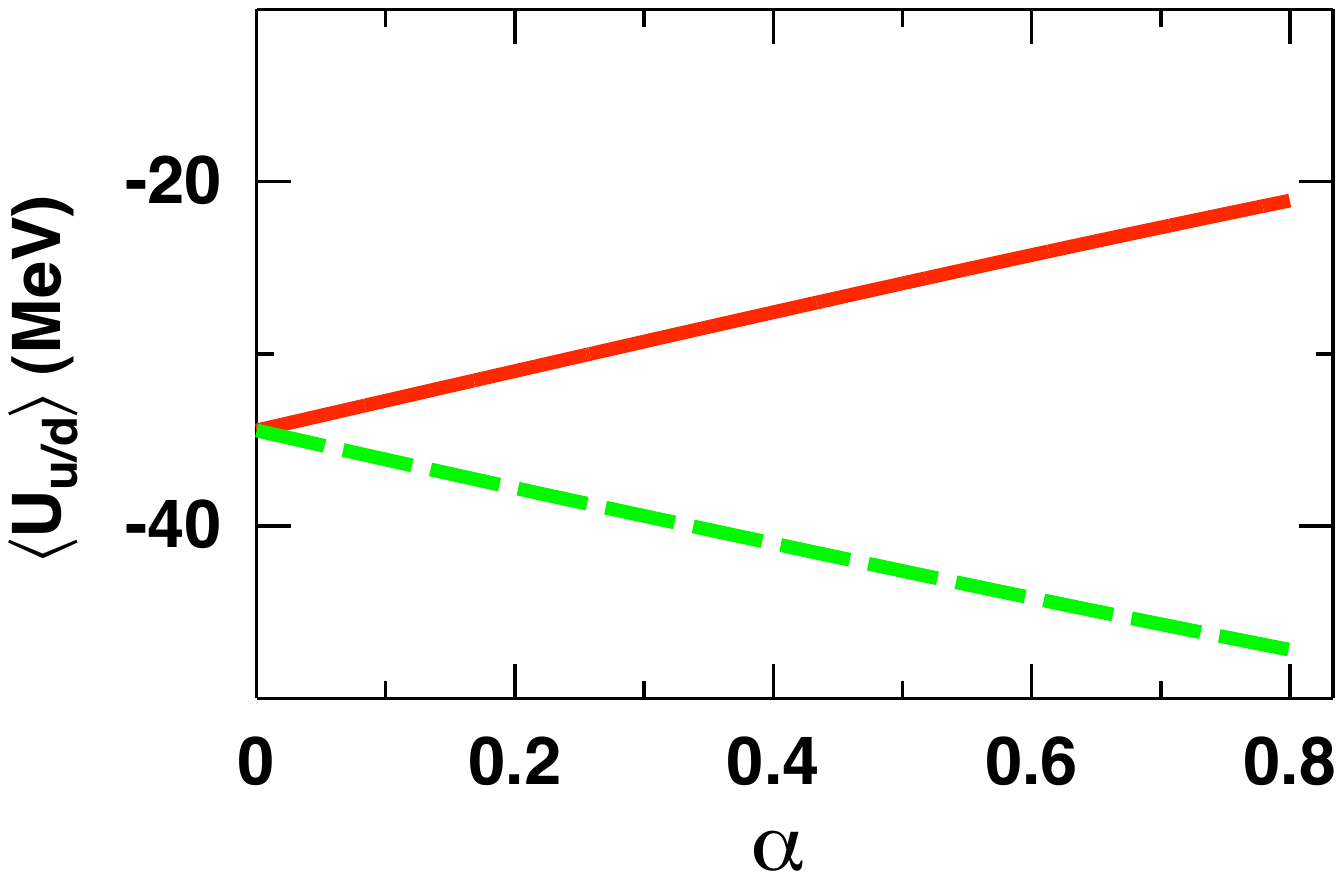}} 
\hspace*{-1.0cm}
\scalebox{0.39}{\includegraphics{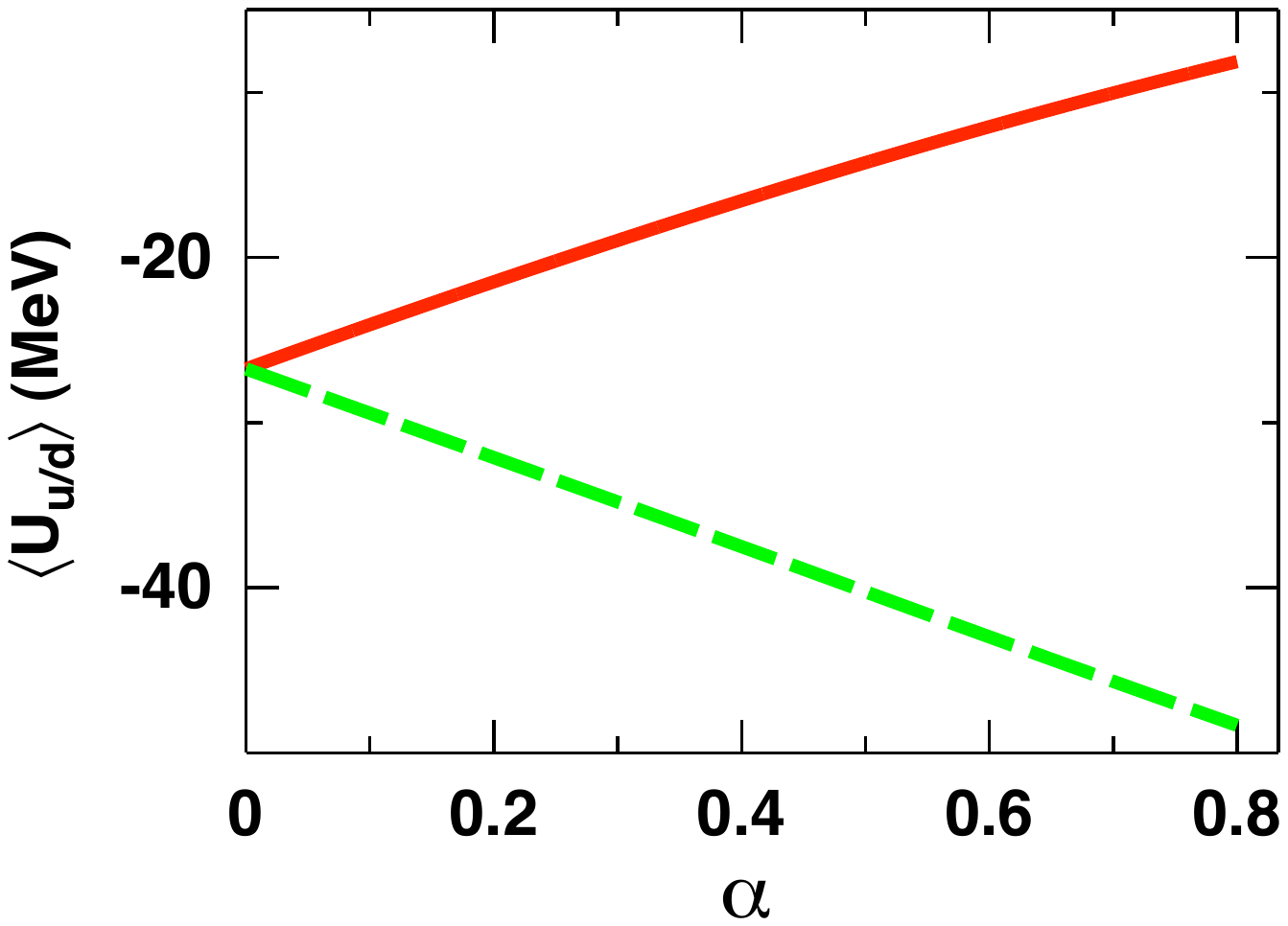}} 
\vspace*{-3.9cm}
\caption{(color online)                                        
The spin splitting of the {\it average} potential energy in polarized nuclear matter as 
a function of the spin asymmetry parameter. The average is taken over three-dimensional momenta. 
The (average) Fermi momentum is equal                                 
1.4$fm^{-1}$ (left frame) and 1.6$fm^{-1}$ (right frame).                                
} 
\label{one}
\end{figure}

\begin{figure}[!t] 
\centering 
\vspace*{-1.0cm}
\hspace*{-2.20cm}
\scalebox{0.45}{\includegraphics{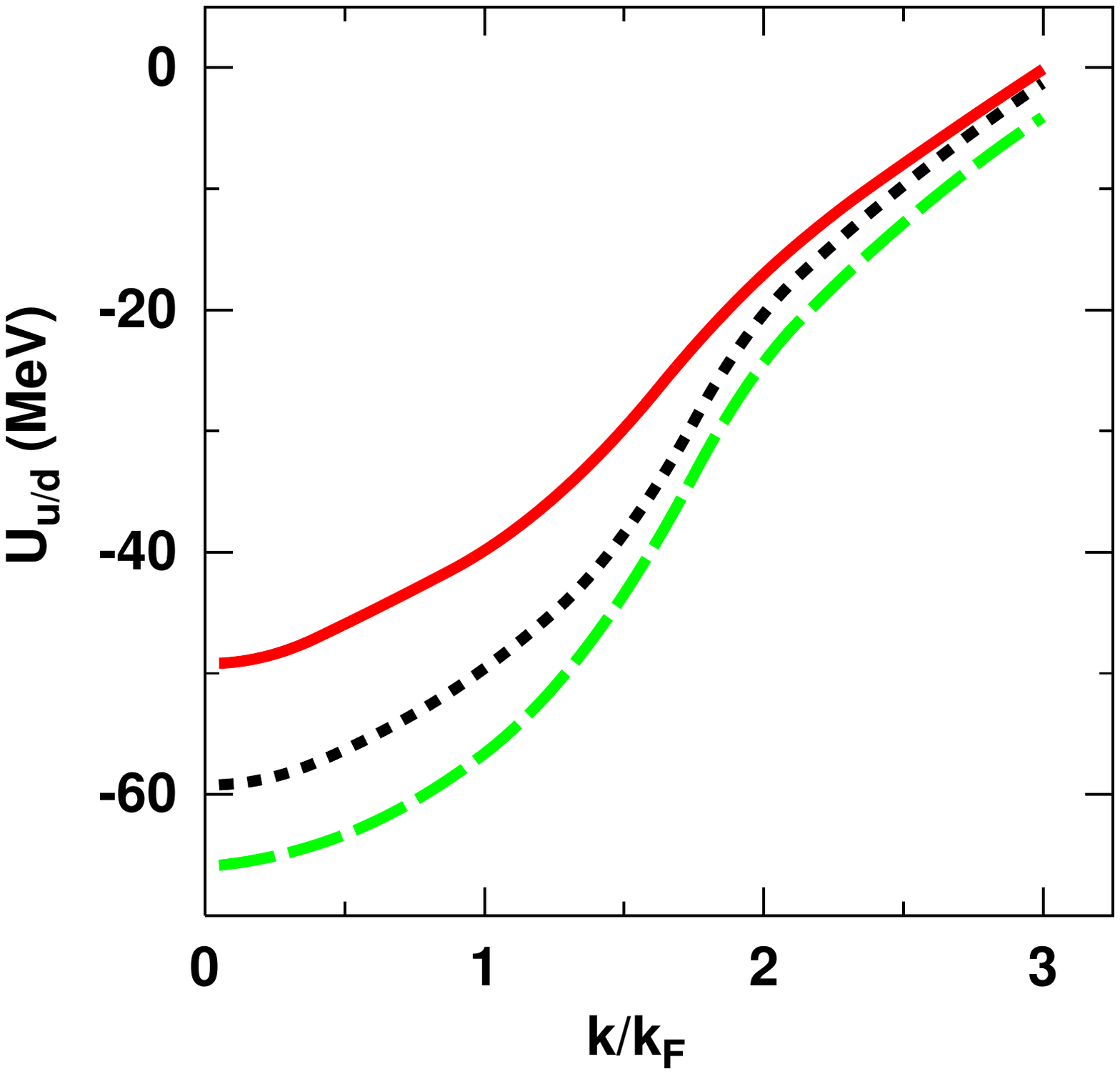}} 
\hspace*{-1.00cm}
\vspace*{1.0cm}
\scalebox{0.45}{\includegraphics{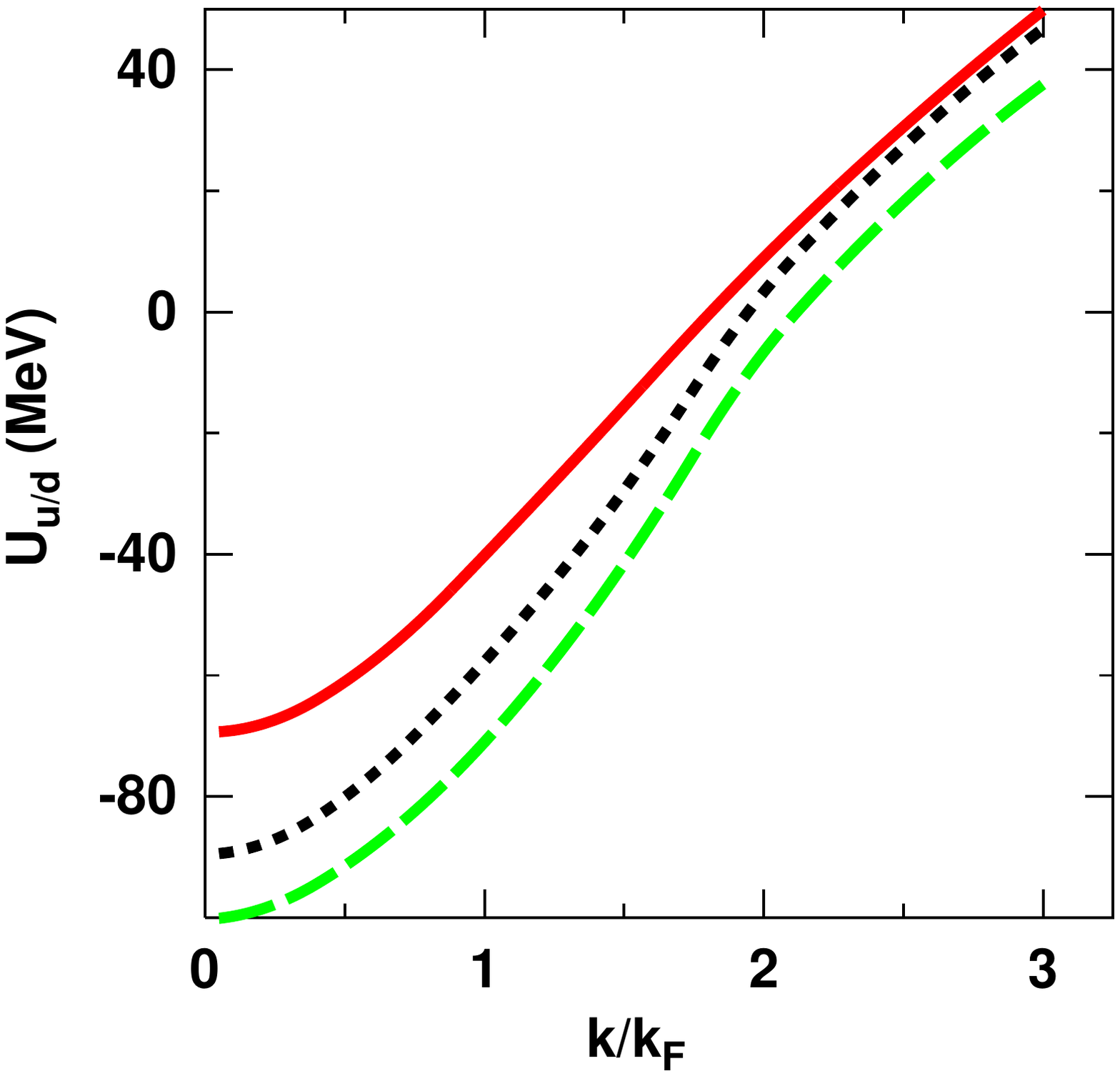}} 
\vspace*{-3.9cm}
\caption{(color online)                                        
The momentum dependence of the single-nucleon potential with spin up (highest curve) and
spin down (lowest curve) in spin-asymmetric matter with $\alpha$=0.6. The middle curve
displays the potential in spin saturated matter. The Fermi momentum is equal to 1.1$fm^{-1}$
in the left frame and 
1.4$fm^{-1}$ in the right frame. 
} 
\label{two}
\end{figure}

\begin{figure}[!t] 
\centering 
\vspace*{-1.0cm}
\hspace*{-1.50cm}
\scalebox{0.50}{\includegraphics{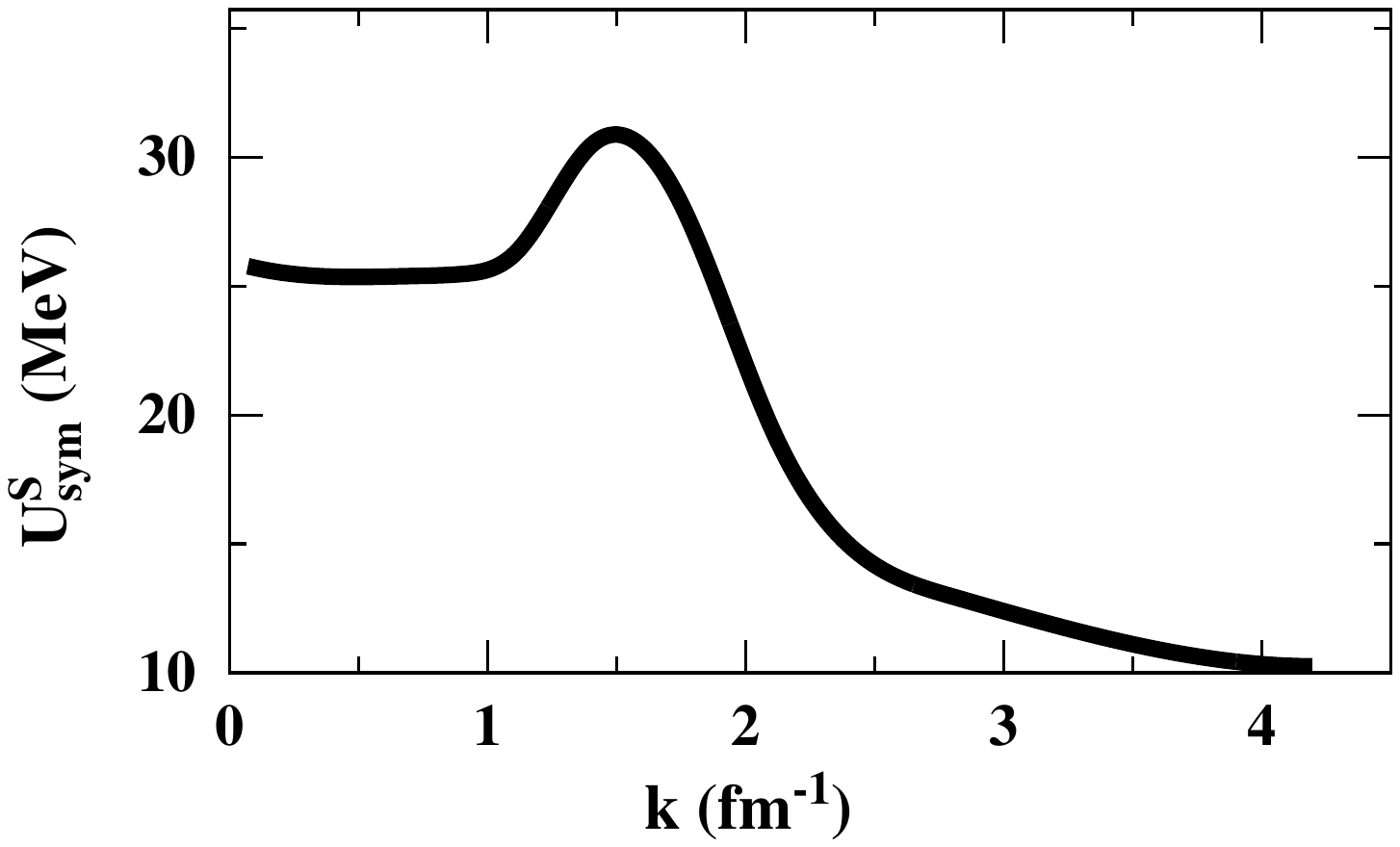}} 
\hspace*{-2.80cm}
\scalebox{0.50}{\includegraphics{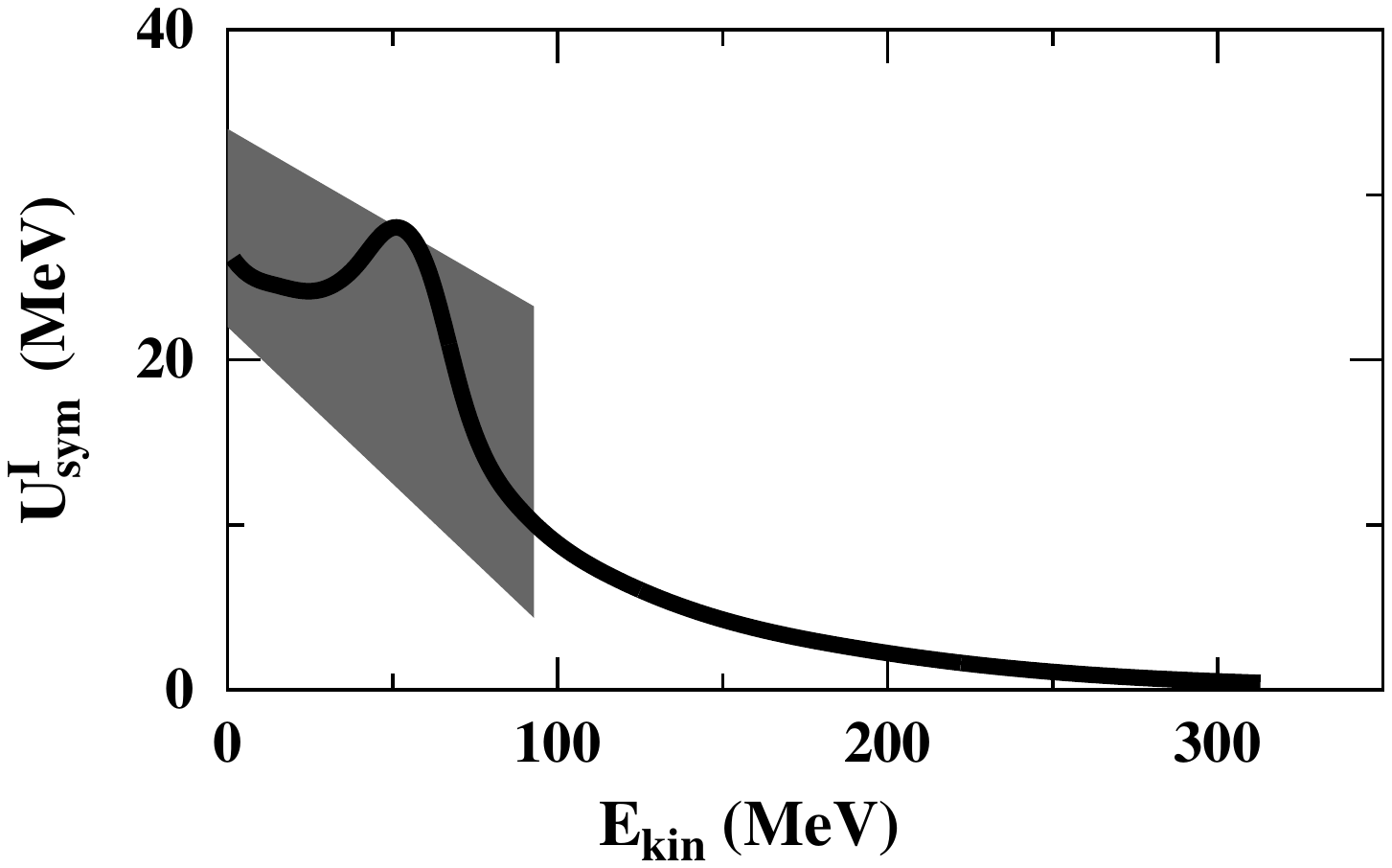}} 
\vspace*{-4.0cm}
\caption{                                                      
Left frame: the spin symmetry potential as a function of the momentum.                
 The Fermi momentum is equal to 1.4$fm^{-1}$. 
The frame on the right shows the symmetry potential in isospin asymmetric (unpolarized)
matter at the same density [33]. The shaded area inside the right panel represents empirical
constraints from Ref.~[30]. 
} 
\label{three}
\end{figure}

From the approximate linear relation apparent from Fig.~1, one can write 
\begin{equation}
<U_{u/d}(\rho,\alpha)> \approx <U(\rho)>_0 \pm <U_{sym}^{S}(\rho)>\alpha  \;,
\end{equation}
where $<U_{sym}^{S}>$ plays the role of an (average) ``spin symmetry" potential, 
\begin{equation}
<U_{sym}^S> = (<U_u>- <U_d>)/2 \alpha \; .                                          
\end{equation}
For a more direct connection with an actual physical experiment, one would write 
(suppressing, for simplicity, $\rho$ and $\alpha$ dependences) 
\begin{equation}
U_{u/d} = U_0 + {\cal{U}}_{\sigma}\frac{{\vec s}\cdot {\vec \Sigma}}{A}      \;,
\end{equation}
where ${\vec s}$ and ${\vec \Sigma}$ are the projectile spin and the expectation value of the target spin operator, respectively, and $A$ is the 
mass number of the target. (The momentum dependence may or may not be taken into account, depending on the
particular analysis.) 
Because
\begin{equation}
\frac{{\vec s}\cdot {\vec \Sigma}}{A} = \frac{1}{A}\frac{1}{2}\sigma_z(\frac{1}{2} N_u-
\frac{1}{2} N_d) \; , 
\end{equation}
and $(N_u -N_d)/A$ is easily identified as the parameter $\alpha$ in the neutron-rich nucleus, one can establish an obvious relation between $\cal{U}_{\sigma}$ of  
Eq.~(8) and $U_{sym}^{S}$ defined as in Eq.~(6) (without average if the momentum dependence is being analyzed). In practice, a spin unsaturated nucleus will also have a net
isospin, which means that $\cal{U}_{\sigma}$, $\cal{U}_{\tau}$, and $\cal{U}_{\sigma \tau}$ would all have to be considered.
Comparison with some older analyses, (based, mostly, on proton scattering on $^{27}$Al and $^{59}$Co and neutron scattering on $^{59}$Co), was performed in Refs.~\cite{Cugn88,Dabr72}. 

Next, we display the momentum dependence of $U_u(k)$ and $U_d(k)$ at some fixed values of $\alpha$ and
for fixed density. The polar angle is also kept fixed (at the value of $\theta$=0) in view of the mild angular dependence mentioned
earlier. 
Again, we see how the spin-up and the spin-down potentials become more repulsive and more attractive,
respectively. It is interesting to analyze the reasons for this behavior, as it sheds light on the similarity 
between spin and isospin asymmetries. First, let us assume, for the sake of simplification, that 
$k_F^u$ is much larger than 
$k_F^d$, so that $U_u$ and $U_d$ get the largest contributions from the $U_{uu}$ and the $U_{du}$ terms, 
respectively (which have the same, larger, integration limit, $k_F^u$). Thus, $U_u - U_d \approx U_{uu}-U_{du}$.
The $ U_{uu}$ term receives contribution only from the $S=1,M_S=\sigma + \sigma' = +1$  matrix elements. 
 Moving on to the $U_{du}$ term, it receives contributions, with equal weights, 
from $S=0,M_S=0$                                                                                 
and from $S=1,M_S=0$  matrix elements.                       
When all of the appropriate weighting factors are taken
into account, the interaction among nucleons with like spin projections turns out to be {\it more repulsive}
than the one among nucleons with opposite spin components. Thus, the scenario becomes analogous to the case
of isospin-asymmetric nuclear matter, where the interaction among like nucleons (with total isospin equal to 1), is more repulsive than the one
among neutrons and protons. 
(It may be useful to mention that all arguments would remain invariant upon exchange of ``u" and 
``d" labels. The physical source of the splitting we observe is in the different nature of the nuclear force
between nucleons with parallel or antiparallel spins.)

The ``spin symmetry potential", $U_{sym}^{S}=(U_u - U_d)/(2 \alpha)$, is displayed in Fig.~3 as a function of 
the momentum. It is remarkably similar, both qualitatively and quantitatively, to the symmetry potential for IANM \cite{FS10} shown on the 
RHS of the figure,                    
$U_{sym}^I=(U_n - U_p)/(2 \alpha)$. Concerning the                                 
latter, it has been shown that, starting from a phenomenological formalism for the single-nucleon potential,
it is possible to predict opposite tendencies for the energy dependence of the symmetry 
potential (while still maintaining nearly the same value of the symmetry energy \cite{Li,Bomb,Rizzo}), resulting in very different predictions of some heavy-ion observables. Similar uncertainties are to be expected with 
$U_{sym}^{S}$.                                                                        

Finally, we 
notice that the approximately linear dependence ({\it vs.} $\alpha$) manifest from Fig.~1, along with a similar 
behavior of the kinetic energy, implies 
the well-known parabolic form for the EoS:
\begin{equation}
<e(\rho,\alpha)> \approx <e_0(\rho)> + <e_{sym}^{S}(\rho)>\alpha^2 \;.
\end{equation}
This is demonstrated in Fig.~4, where the energy/particle (averaged over spin-up and spin-down nucleons) is
compared with the parabolic approximation. 

Before closing, 
 we stress again the importance of more and better empirical constraints to gain insight into the 
spin-dependent part of the nucleon-nucleus optical potential, and, thus, the spin-dependent nuclear 
effective interaction.                        
Valuable information can also come from heavy-ion collisions, provided polarized heavy targets are available.

Within the Landau theory of a Fermi liquid, the effective quasiparticle interaction is represented in terms of functions 
associated with the various 
spin and isospin operators. The Landau parameters are the lowest order terms in the Legendre polynomial
expansions of those functions. 
In particular, the strength of the     
interaction associated with the $\sigma_1\cdot\sigma_2$ operator is represented, to lowest order, by the $g_0$ Migdal-Landau parameter, 
which drives nuclear matter instabilities against spin fluctuations. 
Thus, stringent constraints on the latter, including its density dependence, would 
provide much-needed insight into spin-spin correlations and the possibility of such instabilities.
Because  the expectation value of the $\sigma_1\cdot\sigma_2$ operator is equal to -3 in the singlet states
and +1 in the triplet states, values of $g_0$ which decrease with increasing density 
would signify that the spin-spin force turns less attractive in the singlet configuration and less repulsive
in the triplet one. Thus, there may come a point (in terms of density) when the state with aligned spins is energetically
more favorable than the unpolarized one, resulting in spin instability.

\begin{figure}[!t] 
\centering 
\vspace*{-1.0cm}
\hspace*{-1.0cm}
\scalebox{0.55}{\includegraphics{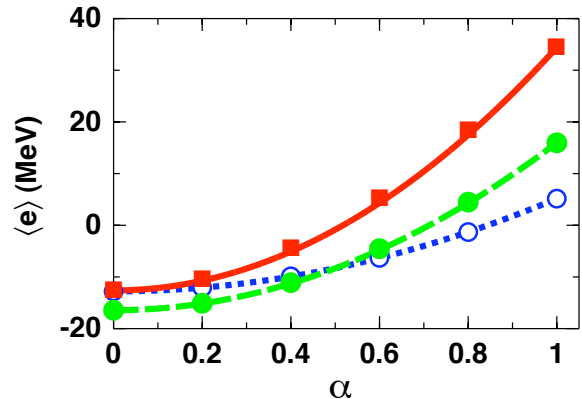}} 
\vspace*{-5.5cm}
\caption{(color online)                                        
The energy/particle in polarized SNM at three fixed densities as a function of the spin asymmetry 
parameter. The solid lines are the parabolic approximation, Eq.~(10), to the calculated values.             
The various densities are in units of $fm^{-3}$. 
} 
\label{four}
\end{figure}

\section{Conclusions}                                                                  

We continue our broad analysis of various phases of nuclear matter. 
Here, we specifically address the splitting of the single-nucleon potential in spin-polarized, but isospin-symmetric, 
nuclear matter. 
The behavior of the predictions is perfectly parallel to the one encountered in 
IANM. 
We point out that additional constraints are crucial for a better understanding of the 
polarizability of nuclear matter. Spin and isospin unsaturated phases of nuclear matter are also 
interesting systems which we plan to study, although computationally more involved. 

As usual, we adopt the microscopic approach for our nuclear matter calculations. 
Concerning our many-body method, we find              
DBHF to be a good starting point to look beyond the normal states of nuclear matter, which  it describes
successfully. The main strength of this method is its inherent ability to effectively incorporate 
crucial TBF contributions through relativistic effects (see Ref.~\cite{FS10} and references 
therein).

\section*{Acknowledgments}
Support from the U.S. Department of Energy under Grant No. DE-FG02-03ER41270 is 
acknowledged.                                                                           

\end{document}